\documentclass[aps,prl,reprint,showpacs,amsmath,longbibliography]{revtex4-1}
\usepackage{amssymb}
\usepackage{mathrsfs}
\usepackage{graphicx}
\usepackage{algorithm}
\usepackage[normalem]{ulem}
\usepackage{color}
\usepackage{bm}
\usepackage{physics}
\usepackage{amsfonts}
\usepackage{mathrsfs}
 \usepackage{algorithmicx}
\usepackage{algpseudocode}

\begin{document}


\title{Unsupervised identification of Floquet topological phase boundaries}



\author{Nannan Ma}
\affiliation{Department of Physics, National University of Singapore, Singapore 117551, Singapore}


\author{Jiangbin Gong}
\email{phygj@nus.edu.sg}
\affiliation{Department of Physics, National University of Singapore, Singapore 117551, Singapore}

\date{\today}

\begin{abstract}

Nonequilibrium topological matter has been a fruitful topic of both theoretical and experimental interest. A great variety of  exotic topological phases unavailable in static systems may emerge under nonequilibrium situations, often challenging our physical intuitions. How to locate the borders between different nonequilibrium topological phases is an important issue to facilitate topological characterization and further understand phase transition behaviors.  In this work, we develop an unsupervised machine-learning protocol to  distinguish between different Floquet (periodically driven) topological phases, by incorporating the system's dynamics within one driving period, adiabatic deformation in the time dimension, plus the system's symmetry all into our machine learning algorithm.    Results from two rich case studies indicate that machine learning is able to reliably reveal intricate topological phase boundaries and can hence be a powerful tool to discover novel topological matter afforded by the time dimension. 

\end{abstract}


\maketitle

\section{Introduction}  

Topological matter is undoubtedly one crucial concept in condensed-matter physics \cite{qi2011topological,hasan2010colloquium} with vast potential. How to generate, distinguish, and characterize different topological phases has led to a number of fascinating research topics.  Considerable theoretical \cite{PhysRevLett.65.3076, Gong2008,PhysRevB.79.081406,kitagawa2010topological,NP2011, PhysRevLett.106.220402,PhysRevLett.109.010601,rudner2013anomalous,asboth2014chiral,nathan2015topological,yao2017topological} and experimental \cite{rechtsman2013photonic,hubener2017creating,mukherjee2020observation,mciver2020light,wintersperger2020realization} efforts have been devoted to studies of  topological phases induced by periodic driving \cite{rudner2020band}, termed as ``Floquet topological phases" (FTPs).  FTPs have fascinating and in many cases exotic features inaccessible from static systems \cite{rudner2013anomalous,tong2013generating,morimoto2017floquet,zhou2018,PhysRevLett.121.036401,zhu2021symmetry,zhu2021floquet}. One seminal example is the anomalous Floquet topological insulator with zero Chern number \cite{rudner2013anomalous} but possessing topological singularity in the time domain. Another example is second-order Floquet topological insulator with zero polarization and vanishing quadrupole moment \cite{zhu2021symmetry,zhu2021floquet}. In such novel phases, the time dimension plays an indispensable and subtle role.
The time dimension inherent to FTPs also makes it challenging to systematically distinguish between different FTPs on a general ground.

In view of the demonstrated capacity of machine learning (ML) \cite{ml1,ml2,ml3,ml4,ml5}, it is a natural attempt to apply ML to distinguish between different topological phases, using either supervised \cite{zhang2018machine,sun2018deep,lian2019machine,holanda2020machine,zhang2021machine} or unsupervised learning \cite{wang2017machine,schafer2019vector,lidiak2020unsupervised}.  In particular, because supervised ML requires a great deal of classified data and hence prior knowledge of the concerned system, unsupervised learning is more appealing when attempting to locate topological phase boundaries.  Previous studies along this avenue have delivered stimulating results on equilibrium systems \cite{long2020unsupervised,balabanov2020unsupervised,scheurer2020unsupervised,che2020topological}. One recent study \cite{PhysRevLett.126.240402} even pushed the unsupervised ML approach to static but non-Hermitian topological matter. However, to date it remains elusive to apply ML to investigate FTPs systematically, with one apparent hurdle being how to correctly incorporate the time dimension into an effective ML algorithm. 

In this work, we demonstrate promising feasibility and effectiveness of ML in locating the phase boundaries between different FTPs. The three key elements in our contribution are (i) to account for system's microscopic motion of the system in the so-called similarity optimization and then the resultant diffusion map \cite{coifman2006diffusion,rodriguez2019identifying} of an unsupervised ML algorithm,  (ii) to consider adiabatic deformation along the time dimension when necessary, and (iii) to specifically exploit various symmetries (if they exist) to simplify the problem.      The first two elements reflect how the time dimension can be universally included in order to identify the FTP boundaries, whereas the third element enhances the notion that symmetry is important in examining different classes of topological phases.

We shall present specific results using two models as case studies, with the second model  hosting a great number of different FTPs including Floquet higher-order phases.   Results represent a highly encouraging advance in ML applications: unsupervised ML can satisfactorily reveal the subtlety of the time dimension  and hence reliably locate intricate topological phase boundaries without calculating any topological invariants.  This work should be of general interest to the identification and characterization of novel FTPs by first predicting rich phase diagrams of nonequilibrium topological matter via ML. 

\section{ML Algorithm}
Consider a periodically driven system whose Hamiltonian satisfies $\textbf{H}(\textbf{\textit{k}},t)=\textbf{H}(\textbf{\textit{k}},t+T)$
with the time variable $t$, the period $T$, driving frequency $\omega=2\pi/T$, and the Bloch wavevector $\textbf{\textit{k}}$.
The time evolution operator of the system for the period from $t_{b}$ to $t_{a}$ can be written as 
\begin{equation}
	U(\textbf{\textit{k}};t_{a},t_{b})=\tau\left\{ \exp [-i\int_{t_b}^{t_a}dt'\textbf{H}(\textbf{\textit{k}},t) ] \right\},
\end{equation}
where $\tau$ represents the time ordering. Throughout we set $ t_b=0 $ without loss of generality and adopt the  simplified notation
$U(\textbf{\textit{k}},t)=U(\textbf{\textit{k}};t_a=t,t_b=0)$.  It is also useful to define an effective Floquet Hamiltonian as follows \cite{nathan2015topological,yao2017topological}:
\begin{align}
	H_{\varepsilon}^{\rm{eff}}(\textbf{\textit{k}})&=\frac{i}{T}\sum_{n}\ln_{-\varepsilon}(U(\textbf{\textit{k}},T))\notag\\
	&=\frac{i}{T}\sum_{n}\ln_{-\varepsilon}[A_n(\textbf{\textit{k}})]\ket{\psi_n(\textbf{\textit{k}})}\bra{\psi_n(\textbf{\textit{k}})},
\end{align}
where  $A_n(\textbf{\textit{k}})$ and $\ket{\psi_n(\textbf{\textit{k}})}$ are respectively the $n$th eigenvalue and eigenvector of the one period time evolution operator $U(\textbf{\textit{k}},T)$. That is, $U(\textbf{\textit{k}},T)=\sum_{n=1}^{N}A_n(\textbf{\textit{k}})\ket{\psi_n(\textbf{\textit{k}})}\bra{\psi_n(\textbf({\textit{k}})})$. Note also that $A_n(\textbf{\textit{k}}) = 1$ or $A_n(\textbf{\textit{k}})=-1$ represents special real eigenvalues on the unit circle, corresponding to Floquet eigenphase 0 or $\pi$. The quasi-energy gaps around  $A_n(\textbf{\textit{k}}) = 1$ and $A_n(\textbf{\textit{k}})=-1$ will be called $0$ gap or $\pi$ gap below.  It is also important to note that the subscript $-\varepsilon$ here specifies the defined branch cut of a logarithm so as to identify which quasi-energy gap we will work with. More specifically,
\begin{equation}
	\ln_{-\varepsilon}\exp(i\phi)=i\phi,\quad    -\varepsilon-2\pi<\phi<-\varepsilon.
\end{equation}
To proceed we further define the following periodic time evolution operator:
\begin{equation}
	U_{\varepsilon}(\textbf{\textit{k}},t)=U(\textbf{\textit{k}},t)\exp[iH_{\varepsilon}^{\rm eff}(\textbf{\textit{k}})t],
\end{equation} 
which clearly satisfies $ U_{\varepsilon}(\textbf{\textit{k}},t)=U_{\varepsilon}(\textbf{\textit{k}},t+T) $.

Suppose now we have a collection of $N$ different systems due to different choices of the system parameters, leading to $N$ different time evolution operators \{$U_{l}({\textbf{\textit{d}}}), l=1,...,N\}$, where \textbf{\textit{d}} stands for (\textbf{\textit{k}}, $t$). From each member of this collection, we define  \{$U_{\epsilon,l}({\textbf{\textit{d}}}), l=1,...,N\}$ as the corresponding unitary periodic in time with period $T$, with $U_{\epsilon,l}\equiv U_{l}({\textbf{\textit{d}}}) \exp[iH_{\varepsilon,l}^{\rm eff}(\textbf{\textit{k}})t]$ and $H_{\varepsilon,l}(\textbf{\textit{k}})$ being the Floquet effective Hamiltonian associated with $U_{l}({\textbf{\textit{d}}})$.  The key question is then the following: Can we design a ML approach to the identification of such $N$ samples into different clusters and thus capture their main differences in terms of their topological features.   This is by construction  a very general question to tackle with, considering that Floquet systems in different parameter regimes may host a great number of topological phases, with differing orders, topological invariants or possible anomalies. 

To consider unsupervised ML, the first step is to quantify the similarity between two different samples. Many quantities may serve this purpose so long as it can roughly depict the strength of connection between different samples. The second necessary step is to introduce the so-called diffusion map \cite{rodriguez2019identifying,coifman2006diffusion} to investigate the similarity distribution before one can classify these samples into different clusters. In essence this is a process to extract the correct clustering information hidden in the similarity distribution.  As seen below, our key contribution is to implement these two steps for FTPs so as to investigate if ML can classify FTPs into different clusters. 

\subsection{Optimization of similarity}

In the first step, we introduce the following quantity to quantify the similarity between sample $l$ and $l^{\prime}$ ($N_{\textbf{\textit{d}}}$ is the number of \textbf{\textit{d}} points in the \textbf{\textit{k}}$-t$ space) when discretizing the space and time variables for the calculation):
\begin{equation}
	S_{l,l'}=\frac{1}{N_{\textbf{\textit{d}}}}\sum_{\textbf{\textit{d}}}\frac{1}{2N_b}{\rm Tr}[U_{\varepsilon,l}^{+}(\textbf{\textit{d}})U_{\varepsilon,l'}(\textbf{\textit{d}})+ h.c],
\end{equation} 
where $ -1 \le S_{l,l'} \le 1, S_{l,l'} =S_{l',l} $, $N_b$ is the number of the bands of the system. By construction $ S_{l,l'} =1 $ if and only if $ U_{\varepsilon,l}(\textbf{\textit{d}})$ and $ U_{\varepsilon,l'}(\textbf{\textit{d}}) $ are identical.  It must be highlighted tha the $t$-dimension and the $k$-dimension are treated above in an equal fashion, thus covering all possible features of the periodic unitary evolution operators within one driving period. 

Whether or not two unitary operators $U_{\epsilon,l}({\textbf{\textit{d}}})$ and $U_{\epsilon,l'}({\textbf{\textit{d}}})$ are topological equivalent is related, but much different from the issue of similarity. From the topological point of view, if they can be deformed to each other smoothly and continuously while preserving their shared symmetries, then they belong to the same topological phase associated with the specific symmetry class.  With this perspective, it is crucial to adopt adiabatic transformation to optimize the similarity between two samples as much as possible, so as to facilitate the identification of topological phase boundaries. For the time evolution unitary operators that form a special unitary (SU) group, such adiabatic deformation can be implemented by considering
\begin{equation}
  S'_{l,l'}=\frac{1}{N_{\textbf{\textit{d}}}}\sum_{\textbf{\textit{d}}}\frac{1}{2N_b}{\rm Tr}[U_{\varepsilon,l}^{\dagger}(\textbf{\textit{d}})\textbf{T}_\textbf{\textit{d}}U_{\varepsilon,l'}(\textbf{\textit{d}})+ h.c],  
\end{equation}
 where the unitary transformations 
 $ \textbf{T}_\textbf{\textit{d}}=\exp{i\sum_{n}\varphi_n(\textbf{\textit{d}})\Lambda_n} $ respect the various symmetries of the system under consideration,  $ \Lambda_n $ is a complete set of hermitian operators to generate all possible deformations, and $ \varphi_n(\textbf{\textit{d}}) $ represents continuous functions of $ \textbf{\textit{d}}$ associated with $U_{\epsilon,l}({\textbf{\textit{d}}})$. That is,
when comparing $U_{\epsilon,l}({\textbf{\textit{d}}})$ and $U_{\epsilon,l'}({\textbf{\textit{d}}})$,
we deform the latter to $\textbf{T}_\textbf{\textit{d}}U_{\varepsilon,l'}
$ and then recompare the deformed unitary with the former. To optimize the similarity between $U_{\epsilon,l}({\textbf{\textit{d}}})$ and $U_{\epsilon,l'}({\textbf{\textit{d}}})$,  a sequence of deformation unitary $ T_\textbf{\textit{d}} $ can be introduced.

\begin{figure*}[htbp]
	\includegraphics[scale=0.33]{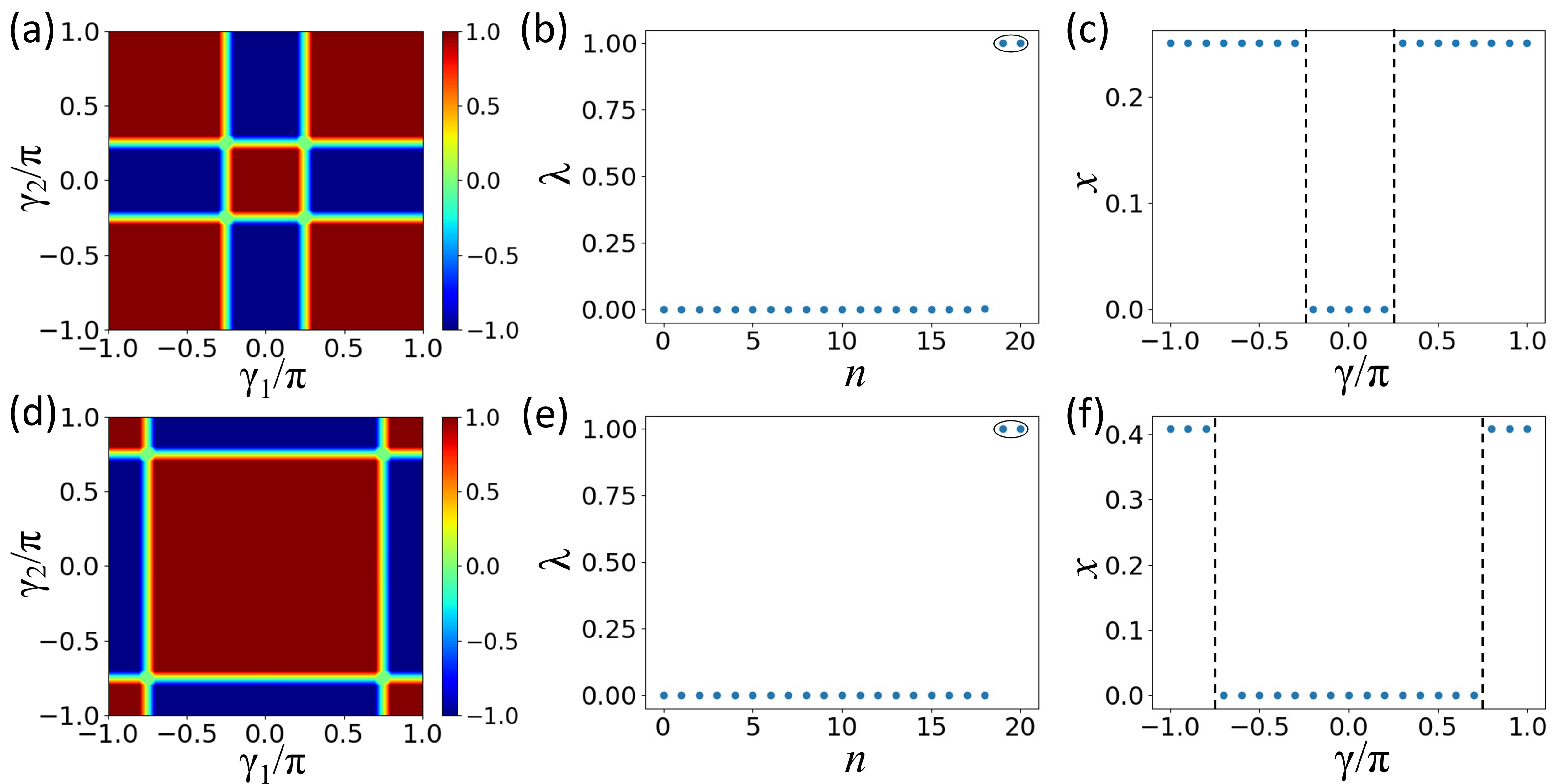}
	\caption{\label{f3} {\bf Detailed results from our ML algorithm using $n=21$ samples  for $\pi$ gap, 0 gap with deformation acting on ``high symmetry" points where $ t=\frac{T}{2} $, with $ \theta=0.75\pi $ and $ \gamma\in [-\pi,\pi] $}.
		({\bf a - c}) Similarity distribution, eigenvalues $\lambda$ of transition matrices and mapping 1D coordinate $x$ of the $ \pi $ gap.  
		({\bf d - f}) Similarity distribution, eigenvalues $\lambda$ of transition matrices and mapping 1D coordinate $x$ of 0 gap. Here, $\eta=0.05, \epsilon=0.03, f=0.5, N_k=20, S_i=0.2, \Lambda_\varphi=0.6$.}
\end{figure*}

The complete set of hermitian generators {$ \Lambda_n $} can be obtained from the dynamical algebra of the periodically  driven system.  For two-band systems used for illustration below, this set is given by Pauli matrices \{$ \sigma_x,\sigma_y,\sigma_z $\}. Next one needs to identify the optimal dependence of $ \varphi_n $ upon \textbf{\textit{d}} with the gradient ascent method: 
\begin{equation}
	\label{e1}
	\varphi_n(\textbf{\textit{d}})	=-\frac{1}{\eta}\frac{1}{N_{\textbf{\textit{d}}}}\frac{1}{2N_b}\rm{Tr}[U_{\varepsilon,l}^{\dagger}(\textbf{\textit{d}})i\Lambda_nU_{\varepsilon,l'}(\textbf{\textit{d}})+ h.c],
\end{equation}
with the learning rate $ \eta \in \mathbb{R}^+ $. Note that in the treatment here, the time dimension is already specifically accounted for because the deformation operators introduced above is allowed to carry an explicit time dependence. Physically speaking, this amounts to moving possible topological singularity along the time dimension in order to extract the best similarity of two Floquet systems.  As another important note, this gradient ascent approach alone cannot guarantee that the deformation on the time evolution operators never changes the topology of the system. To address this potential issue arising from introducing too aggressive deformation, we introduce the overall smoothness function ${Sm}$:
\begin{equation}
	Sm={\rm min}_j\frac{1}{2N_b}{\rm Tr}[U_{\varepsilon,l}^{\dagger}(\textbf{\textit{d}}_j)U_{\varepsilon,l'}(\textbf{\textit{d}}_{j+1})+ h.c],
\end{equation}
where $\textbf{\textit{d}}_j$ are the grid points in the discretized (\textit{\textbf{k}}$-t$) space in our actual calculation, with $ -1 \le Sm \le 1$.  As a practical procedure or rule of thumb we demand that the overall smoothness function before and after the deformation should not drop below a threshold $ \Lambda_s $. That is, 
$S_i-S_d>\Lambda_s$, with $ S_i $ being the initial overall smoothness and $S_d $ the final smoothness after the deformation.   During the deformation process, we halt and set $ S_{l.l'}=-1 $ once the overall smoothness drops beyond $\Lambda_s$.  Furthermore, to ensure that we do not miss some intermediate states $ U_{\varepsilon,l'}^i(\textbf{\textit{d}}) $ whose overall smoothness might be already below the threshold, we require that the deformation in each step is sufficiently small. This motivated us to require that the maximum gradient of $ \varphi_n(\textbf{\textit{d}}) $ is smaller than a cutoff value denoted as $ \Lambda_\varphi $. In practice, both cutoff values of $\Lambda_s$ and $\Lambda_\varphi$ can be set to be rather small when the sampling points in $\textbf{\textit{d}}$ are dense enough. 
In our explicit calculations the values of the cutoff parameters would be regarded as appropriate once the similarity distribution starts to display clear boundaries [e.g., in the case of of Fig. 1(a) and (d) to be discussed below, this refers to some rectangle structures appearing in a 2D parameter space]. 

Given that topological classification itself for equilibrium systems is symmetry based, it can be expected that one must specifically take into account symmetry in the actual ML algorithm. To that end, we denote the symmetry representation of $ g \in \xi $ in space-time \textbf{\textit{d}}  and in the $ N_b $-dimensional vector space respectively as $ \mathcal{R}_d(g) $ and $ \mathcal{R}_u(g) $. In terms of the above-introduced deformation operators, the invariance under $g$ implies   $\varphi_n(\mathcal{R}_d^{-1}(g)\textbf{\textit{d}})\mathcal{R}_u(g)\Lambda\mathcal{R}_u^\dagger(g)=\varphi_n(\textit{\textbf{d}})\Lambda_n$. To guarantee that this is the case, we introduce the following symmetrization \cite{scheurer2020unsupervised}
\begin{equation}
	\label{e2}
	\varphi_n(\textit{\textbf{d}})\Lambda_n \rightarrow \frac{1}{|\xi|}\sum_{g\in\xi}\varphi_n(\mathcal{R}_d^{-1}(g)\textbf{\textit{d}})\mathcal{R}_u(g)\Lambda_n\mathcal{R}_u^\dagger(g).
\end{equation}
Fortunately the previous equation for gradient ascent method is still applicable after this symmetrization.  
Because we treat space and time symmetries under the equal footing, the possible symmetry in the time dimension is also built in. Symmetry consideration may also dramatically speed up or reduce the ML algorithm by reducing the number of sampling points of \textbf{\textit{d}} without costing reliability.  In particular, suppose there are ``high symmetry" points where \textbf{\textit{d}} are invariant upon some symmetry transformation. Then, qualitatively the features of the above-constructed similarity matrix will be dominated by such special points as compared with other points.  This perspective is important to greatly simplify our ML algorithm.
For example, let us assume that the instant Hamiltonian $H(\textit{\textbf{k}},t)$ possesses a chiral symmetry $ S $, that is $S^{-1}H(\textbf{\textit{k}},t)S=-H(\textbf{\textit{k}},-t) $ (which applies to our first case study below). Then the time evolution operator of the system may exhibit the following chiral symmetry $ S^{-1}U_{\varepsilon}(\textbf{\textit{k}},t')S=\pm U_{\varepsilon}(\textbf{\textit{k}},t') $ only at special time points $t'=\frac{T}{2}$ \cite{yao2017topological}. In such cases, a significant reduction emerges if we focus on the \textbf{\textit{k}}-space only and fix the time point at $t=\frac{T}{2}$.  

In the actual execution of the ML algorithm, it is also beneficial to introduce some global deformations to our the samples in order to speed up the identification of similarity \cite{scheurer2020unsupervised}.  To that end, we sample some random, \textbf{\textit{d}}-independent $\varphi_n$ in their certain ranges and inspect the resultant similarity with the target. The transformed one upon the homogeneous unitary transformation that can yield the largest similarity will be used to overwrite the sample unitary we start with.

\subsection{Diffusion map}

With the similarity matrix optimized, we are ready to execute the second step towards locating the FTP  boundaries.  That is, we now adopt the established strategy based on a diffusion map \cite{coifman2006diffusion,rodriguez2019identifying} to amplify the similarities with a Gaussian kernel (referring to one single gap): 
\begin{equation}
	K_{l,l'}^{(1)}=\exp\left[-(1-(\frac{1+S_{l,l'}^{\beta}}{2})^f/\epsilon\right]
\end{equation} 
with $\beta$ referring to either the $0$ or $\pi$ gap under consideration. The amplification coefficients $ f $ and $\epsilon$ are chosen rather arbitrarily in the calculations.  This is analogous to other applications of unsupervised ML, but keeping in mind that the similarity matrix itself has already accounted for the time dimension.  
More importantly, noting that both the $0$ and $\pi$ gap must be considered in order to have a full picture of the phase boundaries.  We hence
introduce another overall two-gap diffusion map
\begin{equation}
K_{l,l'}^{(2)}=\exp\left(-(1-(\frac{1+S_{l,l'}^0}{2}\frac{1+S_{l,l'}^\pi}{2})^f/\epsilon)\right).
\end{equation}
$K_{l,l'}^{(2)}$ will be used below to fully map out the phase diagram in our case studies. 

Based on the matrix $ K_{l,l'}$ (representing either $ K_{l,l'}^{(1)}  $ or$K_{l,l'}^{(2)}$)  with $ l,l'\in [1,N] $,  a fictitious Markov process can be constructed to initiate the diffusion. Such diffusion is described by a one-step transition matrix $P$,  which is defined as:
\begin{align}
	P_{l,l'}=\frac{K_{l,l'}}{z_{l}},\notag\\ z_{l}=\sum_{l'=1}^{N}K_{l,l'},
\end{align}
where $P$ satisfies $ \sum_{l'=1}^{N}P_{l,l'}=1 $.
This way, the similarity matrix is translated to the transition probability of a Markov chain, therefore amplifying the similarity along the diffusion process.
From previous studies it is known that after $2t$ times of transition, the defined diffusion distance matrix $ D_{l,l'} $ is :
\begin{align}
	D_{l,l'}&=\sum_{l''}|(P^t)_{l,l''}-(P^t)_{l',l''}|^2\notag\\
	&=\sum_{k=1}^{N}\lambda_k^{2t}[(\psi_k)_l-(\psi_k)_{l'}]^2 \ge 0.
\end{align} 
Here $ P\psi_k=\lambda_k\psi_k $, $ |\lambda_k|\le 1 $ with $ |\lambda_k|\ge |\lambda_{k+1}| $ and $(\psi_k)_l $ is the $l$th element of the eigenvector $\psi_k$.

Previous work \cite{rodriguez2019identifying} has affirmed that there should be $M$ eigenvalues that are equal to or close to unity if $M$ distinct phases exist.  It is also well known that the eigenvector $\psi_1 $ with the largest eigenvalue is a constant vector whose components are uniformly $ 1/\sqrt{N}$ with $ \lambda_{1}=1$.
As such, one may just focus on $ \psi_k $ where $ |\lambda_k|\simeq 1, k\in[2,M] $ after sufficient time of diffusion has elapsed.  It is then intuitive to regard the vector $ x_l=[(\psi_2)_l,(\psi_3)_l,...,(\psi_M)_l] $ as the coordinate of sample $ l $ in a $((M-1)$-dimensional Euclidean space. The distance between different samples then represents the possible difference of topological origin. Identifying topological phase boundaries is transformed to a clustering problem, which can be solved efficiently use the $k$-means algorithm(Details on this algorithm shown in Supplementary Materials).

\section{Computational Examples}

\subsection{Floquet SSH model}

As a benchmark, we start with a one-dimensional (1D) bipartite model: Floquet Su-Schrieffer–Heeger (SSH) \cite{asboth2014chiral} model with chiral symmetry belonging to the \textbf{A\uppercase\expandafter{\romannumeral3}} class. As shown in Fig.~\ref{f1}(a), the driving consists of two steps of equal duration $\frac{T}{2}$ where $ T=2 $ is the driving period in dimensionless units.  The driving protocol is expressed as the following so that the Hamiltonian possess the chiral symmetry discussed above:
\begin{eqnarray}
	H(k,t)=\begin{cases}
		H_1(k), &0\leq t < \frac{T}{4}\\
		H_2(k), &\frac{T}{4} \leq t < \frac{3T}{4}\\
		H_1(k), & \frac{3T}{4} \leq t < T,
	\end{cases}
\end{eqnarray}
with the time-piecewise Bloch Hamiltonians given by \begin{align}
	H_1(k)&=\theta\sigma_x,\notag\\
	H_2(k)&=\gamma(\cos(k)\sigma_x+\sin(k)\sigma_y),
\end{align}
where $\theta$ represents the strength of intra-cell hopping and $\gamma$ depicts the strength of inter-cell hopping strength, and $ \sigma_{x,y,z} $ are the usual Pauli matrices.
Noting the obvious chiral symmetry $\sigma_zH_i(k)\sigma_z=-H_i(k)$ of the instantaneous Hamiltonian, one identifies the chiral symmetry $ \sigma_zU(k,t)\sigma_z=U(k,-t)$ at arbitrary time. Due to this chiral symmetry, there must be two quasi-energy gaps (0 gap and $\pi$ gap).   To execute the adiabatic deformation while keeping the chiral symmetry, the periodic unitary $U_\epsilon(k,t)$ should obey the corresponding symmetry, $\sigma_zU_\epsilon(k,t)\sigma_z=U_{-\epsilon}(k,-t)\exp(i\frac{2\pi t}{T})$ with $\epsilon \in (0,\pi)$.

\begin{figure*}[htbp]
	\includegraphics[scale=0.33]{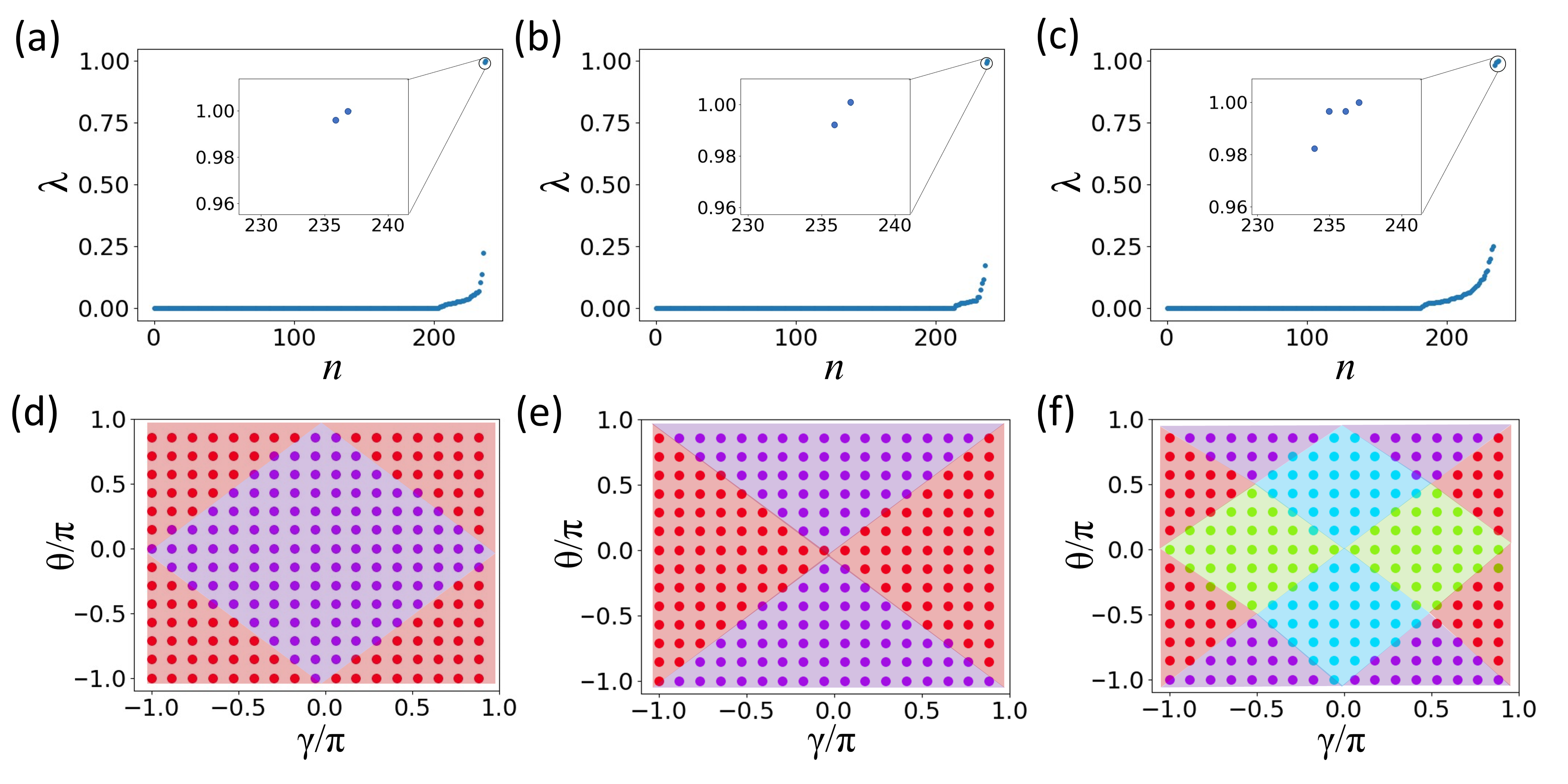}
	\caption{\label{f4} {\bf Eigenvalues $\lambda$ of transition matrices and $K$-mean clustering for $\pi$ gap, 0 gap and the whole system for the Floquet SSH model. Different colors represent different phases. }
		({\bf a - c}) The eigenvalue distributions of transition matrices indicate  that there are two phases in the $\pi$ gap, two phases in the 0 gap and four phases in the whole system.  
		({\bf d - f}) Classification results obtained for $n=17 \times 14$ samples (coloured dots) with $\theta, \gamma \in [-\pi,\pi]$ for $\pi$ gap, 0 gap and the whole system respectively. Different colors represent the associated different phases. Here, $\eta=0.05, \epsilon=0.03, f=0.5, N_t=20, N_k=20, S_i=0.6, \Lambda_\varphi=0.6$.}
\end{figure*}

To execute the similarity optimization step,  the above-described adiabatic deformation step should be in principle applied to the whole $k-t$ space, amounting to a 2-dimensional (2D) parameter space. The symmetrization result is $ \varphi_i(k,t)\sigma_i \to\frac{1}{2}(\varphi_i(k,t)\sigma_i+\varphi_i(k,-t)\sigma_z\sigma_i\sigma_z)$ with $i=x,y,z $.  Results thus obtained with adiabatic deformation acting on the whole $k-t$ space are presented in Supplementary Materials for comparison. Instead we highlight here the results solely using ''high symmetry" points.  As discussed above, the coordinates \textbf{\textit{d}} of high symmetry points do not change under the concerned symmetry transformation. Applying this to the Floquet SSH model here, the special unitaries are $ U_\epsilon(k,T/2)$, thus reducing the optimization to that on the 1D $k$ space.  This reduction speeds up our algorithm and also increases the efficiency of adiabatic deformations. Figure~\ref{f3} depicts the results based on this reduction approach.   As shown in Fig.~\ref{f3}, two clusters can be clearly identified for both $\pi$ gap and 0 gap, by inspecting the similarity distribution. Processing further, the number (denoted $N_c$) of eigenvalues $\lambda_k$ close to unity indicates the number of phases. Fig.~\ref{f3}(b) and (e) present eigenvalues of the transition matrices. Furthermore, low-dimensional mapping space is easy to be visualized and the results are shown in Fig.~\ref{f3}c(f) for the two quasi-energy gaps. In both cases, $N_c=2$ and the phase transition points are identified correctly, at about $\gamma=\pm \frac{\pi}{4}$ for the $\pi$ gap and about $\gamma=\pm \frac{3\pi}{4}$  for 0 gap. It is worth mentioning that the two side regions, namely, $\left|\gamma\right|>\frac{\pi}{4} $ for $\pi$ gap and $\left|\gamma\right|>\frac{3\pi}{4} $ for 0 gap, would be incorrectly identified as the same phase if the adiabatic deformation approach were not used (more details shown in Supplementary Materials).  Compared with the results of Fig.~\ref{f2} obtained considering the whole $k-t$ space, the final classification results
here exploiting ``high symmetry point" $(k,T/2)$ for the 0 and $\pi$ gap indeed suffice and are in agreement with the the results fully sampling the $k-t$ space. There are also some interesting differences.  That is, the similarity distribution associated with the 0 gap has almost no defect by using the ``high symmetry point" $(k,T/2)$; whereas some noise is present when fully sampling the $k-t$ space. The $\pi$ gap case is insensitive to how we perform the sampling.  For these reasons the symmetry-based reduction here actually led to a more clean detection of topologically equivalent phases, thus suggesting that an adiabatic deformation on ''high symmetry" points is an excellent approach in order to capture topologically equivalent phases.

With a significant reduction of computational cost, we are now ready to use ML to explore the whole parameter space $\gamma, \theta\in[-\pi,\pi]$.  The results shown in Fig.~\ref{f4} represent the phase boundaries for $\pi$ gap and 0 gap, in a two-dimensional parameter space. Considering $\pi$ and 0 gap holistically via the two-gap diffusion map, we arrive at the system's overall topological phases, as shown in Fig.~\ref{f3}. Eigenvalues of the transition matrix (Fig.~\ref{f4}(c)) indicate four phases co-existing and the $k$-means method gives a correct phase diagram (Fig.~\ref{f4}(f)). Indeed, for this rather standard model for FTP,  it is not difficult to recognize that this phase diagram contains trivial, 0, $\pi$ and 0$\pi$ phases, corresponding to no edge modes, only 0 edge modes, only $\pi$ edge modes, and coexistence of 0 and $\pi$ edge modes. The overall phase diagram obtained from ML is in remarkable agreement with the true phase plot shown in Fig.~\ref{f1}. 

\subsection{Floquet 2D bipartite model}

Having benchmarked our ML approach using a relatively well understood model in 1D, next we test the capacity of ML applied to
a Floquet two-dimensional (2D) system \cite{zhu2021symmetry,zhu2021floquet}.  The second adopted model accommodates both first-order and higher-order FTPs, and the associated FTP boundaries are rather complicated and not completely understood yet.  The driving protocol acting on a two-dimensional bipartite lattice consists of four steps of  equal duration $T/4$. The respective Hamiltonian as shown in Fig.~\ref {f5} is given by:
\begin{eqnarray}
	H(\textbf{\textit{k}},t)=\begin{cases}
		\theta\sigma_x, & 0\leq t <	\frac{T}{4}\\
		\gamma(e^{ik_y}\sigma^++h.c), &\frac{T}{4} \leq t <\frac{T}{2}\\
		\gamma(e^{i(k_x+k_y)}\sigma^++h.c), &\frac{T}{2} \leq t <\frac{3T}{4}\\
		\gamma(e^{ik_x}\sigma^++h.c), &\frac{3T}{4} \leq t <T\\
	\end{cases}
\end{eqnarray}
where $\theta,\gamma$ represents the strength of hopping between different sublattices and $\sigma^\pm=(\sigma_x\pm\sigma_y)/2$. In our calculations we set $ T=4$ in dimensionless units.  This  model obeys particle-hole symmetry $ C^{-1}H(\textbf{\textit{k}},t)C=-H^*(-\textbf{\textit{k}},t) $ with $ C=\sigma_z $. Due to this feature two quasi-energy gaps around $\varepsilon=0$ or $\varepsilon=\pi$ can be also expected.  Besides, inversion symmetry acts on the lattice as: $ \mathcal{I}H(\textbf{\textit{k}},t)\mathcal{I}=H(-\textbf{\textit{k}},t) $ with $ \mathcal{I}=\sigma_x $.  These two symmetries will have to be fully accounted for in our ML algorithm.   In addition, from previous work \cite{ zhu2021symmetry,zhu2021floquet} it is known that  this model accommodates anomalous Floquet high-order (AFHO) phases in addition to anomalous chiral edge states. All these present a very high-pressure test for our advocated ML algorithm.  

The symmetrization transformation we carried out is the following: $ \varphi_i(k,t)\sigma_i \to\frac{1}{4}(\varphi_i(k,t)\sigma_i-\varphi_i(-k,t)\sigma_z\sigma_i^*\sigma_z+\varphi_i(-k,t)\sigma_x\sigma_i\sigma_x-\varphi_i(k,t)\sigma_y\sigma_i^*\sigma_y) $ with $i=x,y,z$. Note that either particle-hole symmetry or inversion symmetry changes $(\textbf{\textit{k}},t)$ to $(-\textbf{\textit{k}},t)$. Learning from our previous example, this time we take four high-symmetry points $\textbf{\textit{k}}_{hs}$ in the $k$ space, namely,  
$k_x=[0,\pi] $ and $ k_y=[0,\pi]$, in order to most efficiently carry out our ML algorithm. The symmetrization then reduces the adiabatic deformation operators to $ T(\textbf{\textit{k}}_{hs},t)=e^{i\varphi(\textbf{\textit{k}}_{hs},t)\sigma_x} $. 

\begin{figure*}[htbp]
	\includegraphics[scale=0.33]{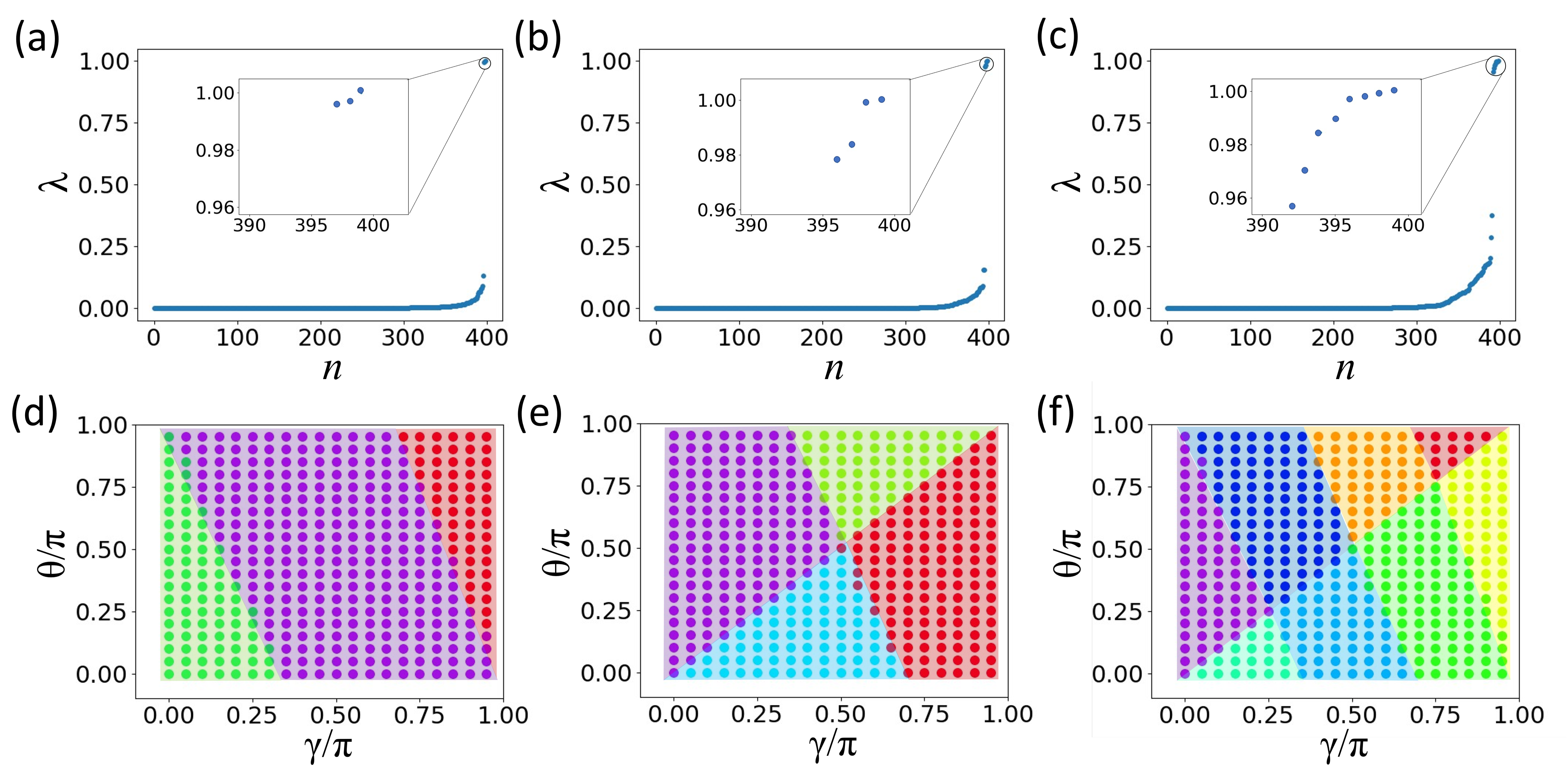}
	\caption{\label{f6} {\bf Eigenvalues of transition matrices and k-mean clustering for $\pi$ gap, 0 gap and the whole system for Floquet 2D bipartite model. Different colors represent different phases by inspecting their mutual distances based on the $k$-means algorithm.}
		({\bf a - c}) Eigenvalue distributions of transition matrices indicate  that there are three phases in the $\pi$ gap, four phases in the 0 gap and eight phases in the whole system.
		({\bf d - f}) Phase boundaries obtained for $n=20 \times 20$ samples (coloured dots) with $\theta, \gamma \in [-\pi,\pi]$ for $\pi$ gap, 0 gap and the whole system respectively, with different colors representing the associated different phases. Here, $\eta=0.5, \epsilon=0.03, f=0.5, N_t=20, S_i=0.2, \Lambda_\varphi=0.6$.}
\end{figure*}

We now investigate the system in two quasi-energy gaps, covering the parameter ranges $ \theta,\gamma \in (0,\pi) $, with the main results presented in Fig.~\ref{f6}.  Three eigenvalues close to unity in Fig.~\ref{f6}(a) indicate three phases existing in the $\pi$ gap.  Specifically,  two phase transition lines, $ \theta+3\gamma=\pi $ and $ \theta+3\gamma=3\pi $ are detected in this phase diagram. Similarly, there are four different phases regarding the 0 quasi-energy gap, as indicated in Fig.~\ref{f6}(b). The final phase diagram reveals two phase transition lines for the 0 gap, $ \theta=\gamma $ and $ \theta+3\gamma=2\pi $. Overall, considering the topology with respect to both quasi-energy gaps, we obtain Fig.~\ref{f6}(c,f) that gives a total of eight FTPs, with the obtained phase boundaries in excellent agreement with what was previously found from this very rich system (see Fig.~\ref{f5}). For example, upon further analysis under the open boundary condition, the regime marked by yellow in Fig.~\ref{f6}(f) turns out to be an anomalous Floquet second-order topological insulator phase. The orange and light blue regions correspond to two different anomalous Floquet topological insulator phases with chiral edge states existing in both gaps. The phase represented by purple triangles is a trivial phase without any topological boundary states. The other four regions are different Chern insulators with chiral edge states in just one gap which can be described by the Floquet effective Hamiltonian. The possibility to be able to distinguish between such FTPs including exotic and anomalous phases by ML is both stimulating and promising. It should be highlighted that this intricate phase diagram is obtained without requiring prior knowledge about bulk topological invariants, bulk-edge correspondence, dynamical reduction etc. 

\section{Conclusion}

We have proposed an unsupervised ML algorithm to locate the possibly intricate boundaries between different FTPs, without prior knowledge of how to explicitly characterize the various topological phases or using any result under the open boundary condition. Our approach is strongly supported by two case studies. Central to our ML algorithm, we have incorporated the concept of adiabatic deformation along the time dimension and thus naturally and effectively  capture possible topological singularities in the time domain that cannot removed unless a topological transition occurs. Though not necessary in principle, we have also cleverly used the high-symmetry points to reduce the computation of the similarity matrix.  This strengthens the known importance of symmetry analysis in topological classification, but now purely from a computational point of view.  
What is particularly intriguing is that our ML algorithm can even locate the boundary between first-order FTPs and second-order FTPs. 

Our success here represents another fascinating application of ML. On the one hand, artificial intelligence is able to guide us to design optimized driving protocols to achieve desired topological features such as very large topological invariants \cite{Zhang2019}; on the other hand we now further show that ML can be used to map out highly nontrivial phase diagrams before detailed topological characterization can be done.   We hence conclude that our ML algorithm incorporating the time dimension can be a powerful tool towards the discovery of more exotic FTPs, synthesized by engineered control fields.   For example, it can be highly useful in treating non-Hermitian Floquet topological phases, which are subtle due to the interplay of non-Hermitian effects and the effective long-range hopping due to periodic driving \cite{Zhang2020,Zhou2021}. 

\begin{acknowledgements}
We acknowledges funding support by the Singapore NRF Grant No. NRF-NRFI2017-04 (WBS No. R-144-000-378- 281).
\end{acknowledgements}

\appendix
\begin{widetext}
\section{SUPPLEMENTARY MATERIALS}
\renewcommand{\theequation}{S.\arabic{equation}}
 \setcounter{figure}{0} 
\renewcommand{\thefigure}{S\arabic{figure}}
\setcounter{equation}{0}

\subsection{k-means algorithm}
In the context of unsupervised learning, clustering is a general task for processing unknown data. It is used to classify samples in the data set into several disjoint sets which are called as clusters. If all samples have been mapped into an Euclidean space, then  the measure is just the Euclidean distance.  In our case, the $k$-means algorithm introduced below is chosen to preform this clustering task.

As the result of the diffusion map, we get a sample set $D=\{\textbf{\textit{x}}_1,\textbf{\textit{x}}_2,...,\textbf{\textit{x}}_N\}$ and the eigenvalues of the transition matrices indicate $k$ disjoint clusters $C=\{C_1,C_2,...,C_k\}$(in our case, $k=M$). Taking into account that the distance $\left|\textbf{\textit{x}}_l-\textbf{\textit{x}}_{l'}\right|^2$ quantifies the similarity between sample $l$ and $l'$, $k$-means attempts to minimize the distance square of all clusters:
\begin{equation}
    E=\sum_{i=1}^k\sum_{\textbf{\textit{x}} \in C_{i}}\left|\textbf{\textit{x}}-\boldsymbol{\mu}_{i} \right|^2,
\end{equation}
where $\boldsymbol{\mu}_i=\frac{1}{\left|C_i\right|}\sum_{\textbf{\textit{x}} \in C_i}\textbf{\textit{x}}$ is the mean vector of cluster $C_i$.

In principle, finding the optimal solution requires to check all possible cluster classifications. It is not an easy task. The $k$-means algorithm makes use of iterative optimization with greedy policy to approach the idea solution. The details of algorithm flow are shown in the following table ~\ref{alg:the_alg}. The first line initializes mean vectors, lines four to eight and lines nine to sixteen are iterative updates of cluster classification and mean vectors respectively, then return the current classification at line eighteen  if it is unchanged after an iteration. This way we can obtain the right result with a high probability.
In our case studies, these clusters represent different topological phases. This means that samples belonging to the same cluster have the same topological phase. We then assign different colors to different clusters, from which we can obtain clear phase diagrams. 

\begin{algorithm}[H]
  \caption{ k-means.}  
  \label{alg:the_alg}
  \begin{algorithmic} [1]
    \Require  
      Sample set $D=\{\textbf{\textit{x}}_1,\textbf{\textit{x}}_2,...,\textbf{\textit{x}}_N\}$  
    \Ensure  
      Clusters $C=\{C_1,C_2,...,C_k\}$        
    \State  Select k samples randomly as the initial mean vectors$\{\boldsymbol{\mu}_1,\boldsymbol{\mu}_2,...,\boldsymbol{\mu}_k\}$ from $D$
    \Repeat
        \State Set $C_i=\varnothing(1\le i \le k)$
        \For{$j=1,2,...,N$}
            \State Calculate the distances between sample $\textbf{\textit{x}}_j$ and each mean vector $\boldsymbol{\mu}_i(1\le i \le k)$: $d_{ji}=\left|\textbf{\textit{x}}_j-\boldsymbol{\mu}_{i} \right|^2$;
            \State Identify the cluster mark $m_j$ of sample $\textbf{\textit{x}}_j$ according to the shortest distance: $m_j=arg\,min_{i \in \{1,2,...,k\}}d_{ji}$;
            \State Put sample $\textbf{\textit{x}}_j$ in respective cluster:
            $C_{m_j}=C_{m_j}\bigcup\{\textbf{\textit{x}}_j\}$;
        \EndFor
        \For{$i=1,2,...,k$}
            \State Calculate new mean vectors: $\boldsymbol{\mu}_i'=\frac{1}{C_i}\sum_{\textbf{\textit{x}}\in C_i}\textbf{\textit{x}}$;
            \If{$\boldsymbol{\mu}_i'=\boldsymbol{\mu}_i$}
                \State $\boldsymbol{\mu}_i$ is updated to $\boldsymbol{\mu}_i'$
            \Else 
                \State Leave the current mean vector $\boldsymbol{\mu}_i$ unchanged
            \EndIf
        \EndFor
    \Until{All current mean vectors are unchanged}\\
    \Return{Final $C=\{C_1,C_2,...,C_k\}$}  
  \end{algorithmic}
\end{algorithm}

\subsection{Models and exact phase diagrams in 1D and 2D Floquet bipartite models}
 The driving protocol and the associated phase diagram of the first Floquet 1D SSH system we studied in the main text are shown in Fig.~\ref{f1}.  The phase diagram consists of four phases, trivial phase, 0 phase, $\pi$ phase and 0$\pi$ phase, possessing no edge states, one type of edge states with zero eigenphase, one type of edge states with $\pi$ eigenphase, and two different types of edge states coexisting, with 0 and $\pi$ eigenphases.

\begin{figure}[htb]
	\includegraphics[scale=0.55]{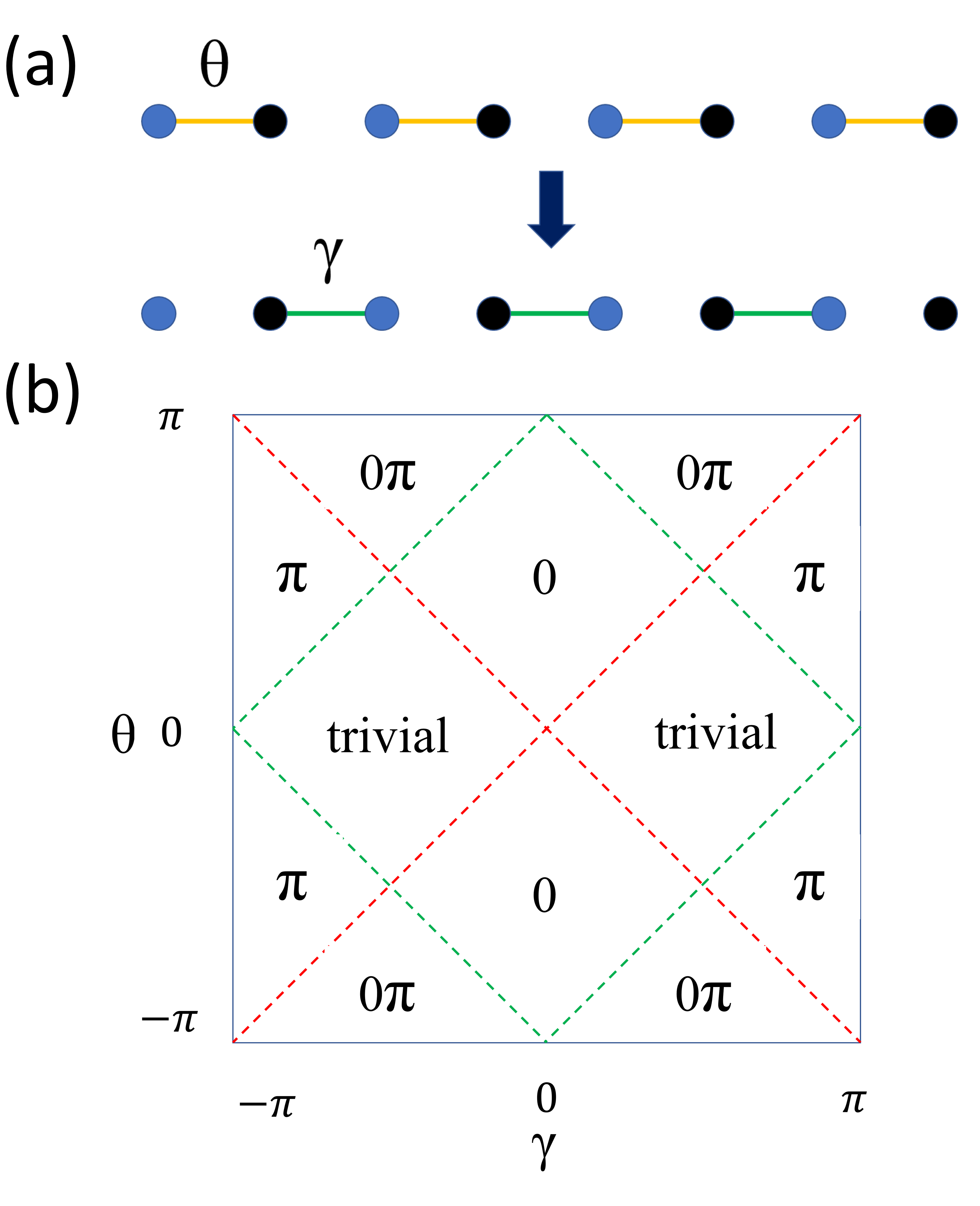}
	\caption{\label{f1} {\bf Periodic modulation protocol and phase diagram of the Floquet SSH model}
		({\bf a}) Driving protocol in a 1D bipartite chain. In each time step, there is either inter-cell hopping with hopping strength $\gamma$ or intra-cell hopping with hopping strength $\theta$. 
		({\bf b}) Phase diagram (obtained by counting certain winding numbers as in previous work) depicts four phases: trivial phase, and phases with one pair of 0 edge modes, one pair of $\pi$ edge modes and with coexistence of 0 and $\pi$ edge mode, with $\gamma, \theta \in [-\pi,\pi]$. The green(red) lines represent the phase transition at $ \pi$(0) gap. As shown in the main text, our ML algorithm can easily reproduce this phase diagram.}
\end{figure}

The driving protocol and the phase diagram of the second Floquet 2D bipartite model are depicted in Fig.~\ref{f5}. The complicated phase diagram can be obtained from considerations of multiple topological invariants along with dynamical reduction, states under open boundary condition, and some symmetry analysis when necessary. 
Based on detailed physical analysis under both periodic and open boundary conditions, abundant FTPs were previously found: one normal insulator (NI) phase without topological boundary states; four different Chern insulators (CI) with chiral edge states occurring in one gap; two anomalous Floquet topological insulator (AFI) where both gaps host chiral edge states and one anomalous Floquet high-order topological insulator (AFHOTI) phase with corner states existing in both gaps. 

\begin{figure}[htbp]
	\includegraphics[scale=0.33]{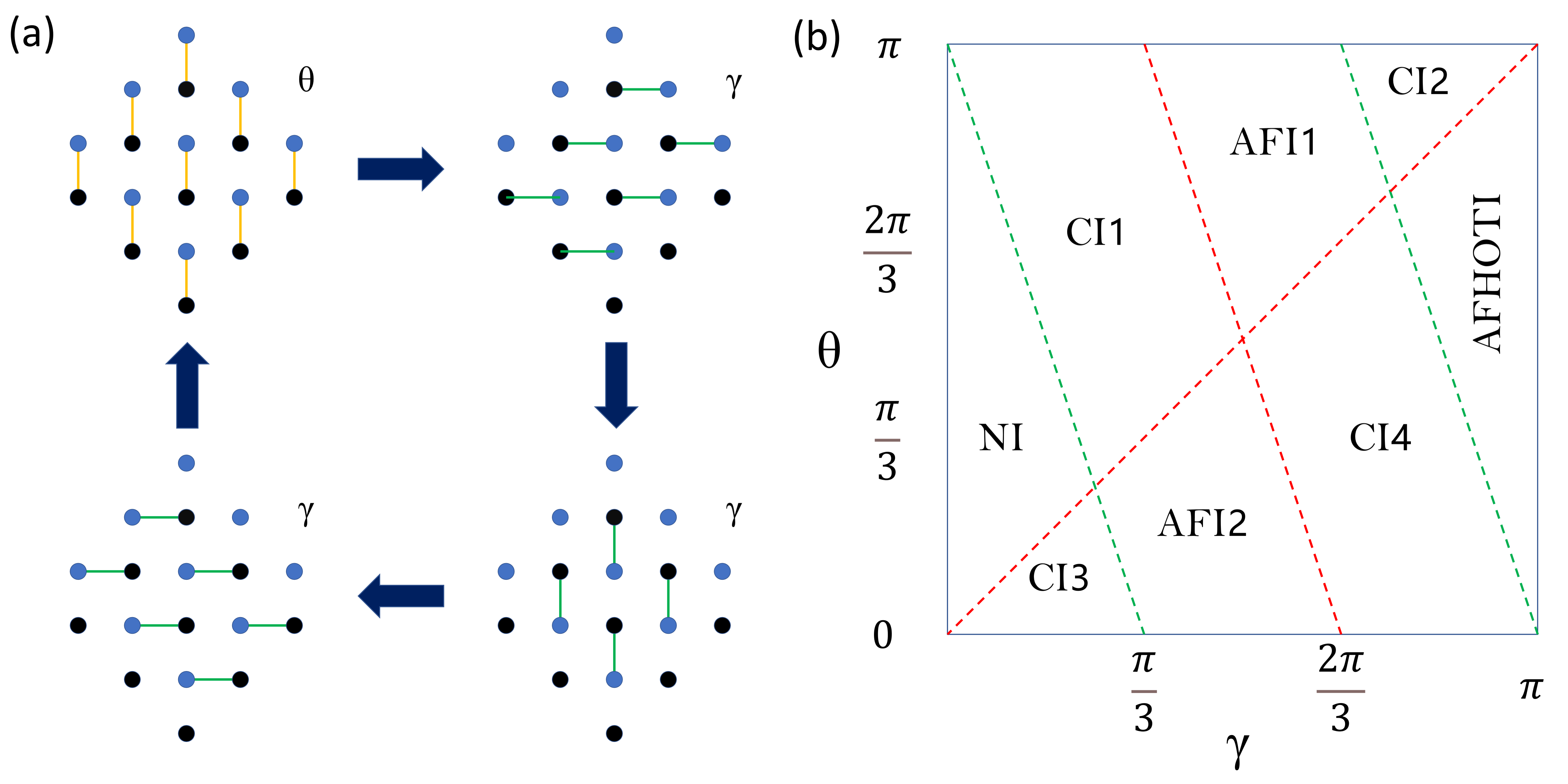}
	\caption{\label{f5} {\bf Periodic modulation protocol and phase diagram of Floquet 2D bipartite model}
		({\bf a}) Driving protocol for sites in 2D bipartite lattice. There are four steps consisting of inter-cell coupling $\gamma$ or intra-cell coupling $\theta$ in each step. 
		({\bf b}) The phase diagram of  this model, containing eight phases with $\gamma, \theta \in [-\pi,\pi]$. The green (red) lines represent the phase transition lines for the $ \pi$(0) gap.}
\end{figure}

\subsection{The result of Floquet SSH model with adiabatic deformation acting on the whole k-t space}

In the main text, it is highlighted that the use of ``high symmetry" points plays a significant role in reducing computational cost without losing accuracy.
Let us now confirm this by presenting analogous results with adiabatic deformation acting on the whole $k-t$ space. In Fig.\ref{f2}, we illustrate the associated similarity distribution  with fixed $\theta = \frac{3\pi}{4}$. Compared to the similarity distribution of just considering ''high symmetry" points in Fig.~\ref{f3}, it is not surprising that the distributions Fig.~\ref{f2}a,b are not perfect as viewed from the results based on the ''high symmetry" result (Fig.~\ref{f3}).  This is also expected due to a significant increase in computational complexity. However, as shown in Fig.~\ref{f2}(b) and (c), this ``rough" similarity matrix is also good enough for diffusion map to identify two phases. The final mappings (Fig.~\ref{f2}c,d) in the 1D $\gamma$-space classify these samples into two phases correctly. The results here hence verify that our reduction of computation by just considering the ''high symmetry" point is valid.  We can also conclude that similarity detection may become more accurate as a result of the decrease in computational complexity by considering ''high symmetry" points.

\begin{figure*}[htbp]
	\includegraphics[scale=0.33]{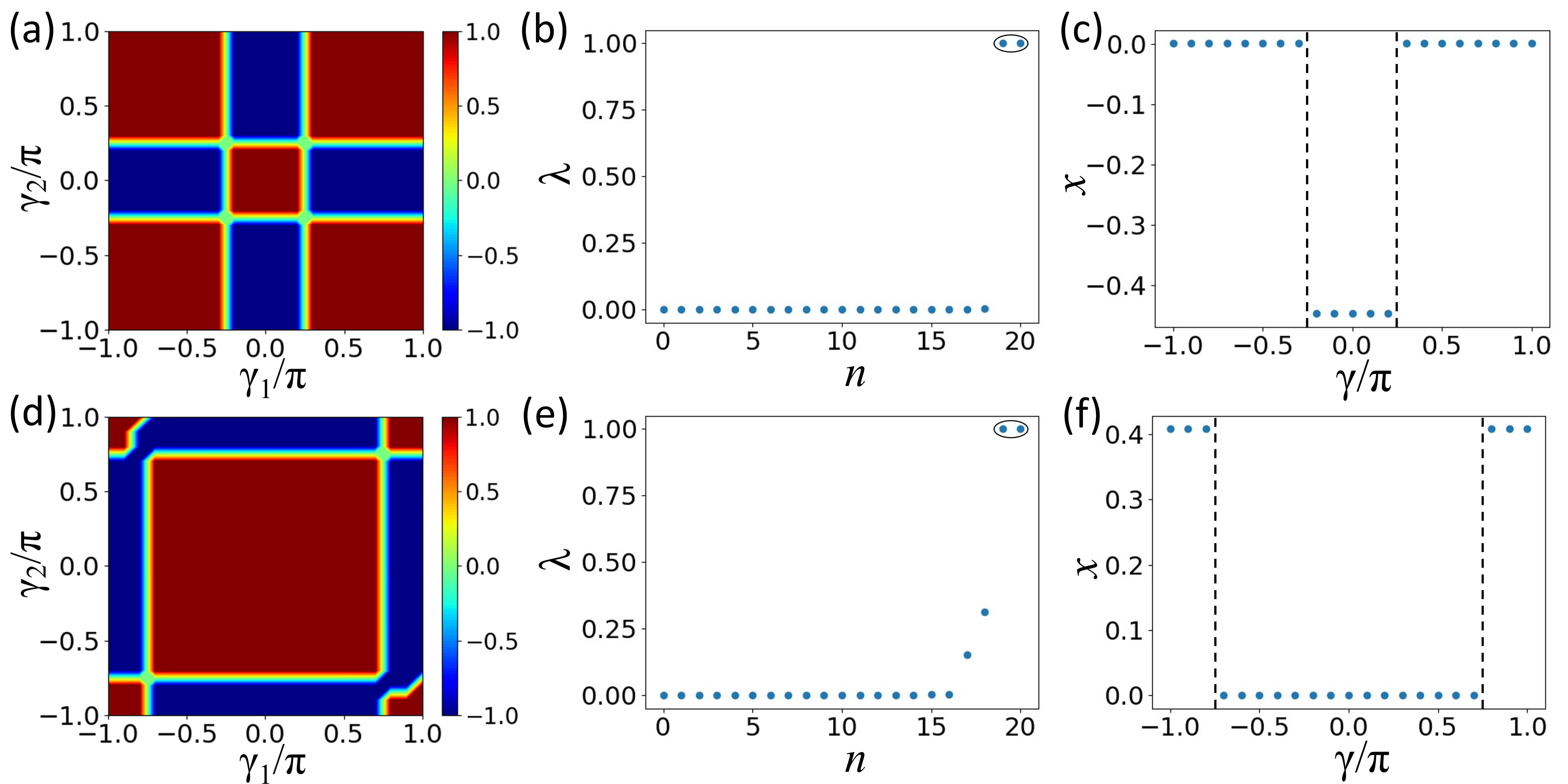}
	\caption{\label{f2} {\bf The results of $n=21$ samples  for $\pi$ gap, 0 gap with deformation acting on whole $k-t$ space when $ \theta=0.75\pi $ and $ \gamma\in [-\pi,\pi] $}.
		({\bf a - c}) Similarity distribution, eigenvalues $\lambda$ of transition matrices and mapping 1D coordinate $x$ for the $ \pi $ gap 
		({\bf d - f})Similarity distribution, eigenvalues $\lambda$ of transition matrices and mapping 1D coordinate $x$ for the 0 gap. Here, $\eta=0.01, \epsilon=0.03, f=0.5, N_t=10, N_k=20, S_i=0.2, \Lambda_\varphi=0.6$.}
\end{figure*}

\subsection{An intuitive explanation of the similarity}

Here we use the Floquet 2D bipartite system to shed more light on our measure of similarity between two unitary operators.

The Brillouin zone $(\textbf{\textit{k}},t)$ is a $I_3$ space and the time evolution operators constitute $ SU(2) $ group. There is a mapping that maps a $ SU(2) $ element to a point on a sphere with a $\pi$ radius, with the center corresponding to identity $ E $ and the surface is $ -E $.  A time evolution operator can be written as: $ U=e^{i\alpha\vec{n}\cdot\vec{\sigma}} $ where $ \vec{n} $ is the direction vector and $ \alpha $ is the length. Then, the similarity $ s_{1,2} $ between $U_1$ and $U_2$ is:
\begin{align}
	s_{1,2}&=\frac{1}{4}\rm{Tr}[U_1^\dagger U_2+h.c]\notag\\
	&=\frac{1}{4}Tr[(\cos\alpha_1-i\sin\alpha_1\vec{n}_1\cdot\vec{\sigma})(\cos\alpha_2-i\sin\alpha_2\vec{n}_2\cdot\vec{\sigma})+h.c]\notag\\
	&=\cos\alpha_1\cos\alpha_2+\sin\alpha_1\sin\alpha_2\vec{n}_1\cdot\vec{n}_2.
\end{align}
with $ -1 \leq s_{1,2} \leq 1 $ and $ s_{1,2}=1 $ if and only if $ U_1=U_2 $. This formula takes into account the direction and length by dot product and trigonometric function.

Then, our similarity formula can quantify the geometrical differences between the mappings of different systems properly. 

\subsection{Gradient of adiabatic deformation under the symmetrization}

Here, we will prove explicitly that the Eq.~\eqref{e1} does not change under the symmetrization (Eq.~\eqref{e2}).

After $ U_{\varepsilon,l'}(\textbf{\textit{d}}) \to U_{\varepsilon,l'}(\textbf{\textit{d}}) $, we obtain a different similarity:
\begin{equation}
	S_{l,l'}=\frac{1}{N_{\textbf{\textit{d}}}}\sum_{\textbf{\textit{d}}}\frac{1}{2N_b}Tr[U_{\varepsilon,l}^{\dagger}(\textbf{\textit{d}})e^{i\frac{1}{|\xi|}\sum_{g\in\xi}\varphi_n(\mathcal{R}_d^{-1}(g)\textbf{\textit{d}})\mathcal{R}_u(g)\Lambda_n\mathcal{R}_u^\dagger(g)}U_{\varepsilon,l'}(\textbf{\textit{d}})+ h.c].
\end{equation}
Here, we set $\textbf{\textit{d}}_0$ as a point in $ \textbf{\textit{d}} $ space and $\textbf{\textit{d}}'$ are points which satisfy $ \mathcal{R}_d^{-1}(g)\textbf{\textit{d}}'=\textbf{\textit{d}}_0 $. The gradient at $\textbf{\textit{d}}_0 $ is:
\begin{align}
    \label{s3}
	\frac{\partial S_{l,l'}}{\partial \varphi_n(\textbf{\textit{d}}_0)}&=\frac{1}{N_{\textbf{\textit{d}}}}\frac{1}{|\xi|}\sum_{\textbf{\textit{d}}'}\frac{1}{2N_b}Tr[U_{\varepsilon,l}^{\dagger}(\textbf{\textit{d}}')i\mathcal{R}_u(g)\Lambda_n\mathcal{R}_u^\dagger(g)U_{\varepsilon,l'}(\textbf{\textit{d}}')+ h.c]  \notag\\
	&=\frac{1}{N_{\textbf{\textit{d}}}}\frac{1}{|\xi|}\sum_{\textbf{\textit{d}}'}\frac{1}{2N_b}Tr[\mathcal{R}_u^\dagger(g)U_{\varepsilon,l}^{\dagger}(\textbf{\textit{d}}')i\mathcal{R}_u(g)\Lambda_n\mathcal{R}_u^\dagger(g)U_{\varepsilon,l'}(\textbf{\textit{d}}')\mathcal{R}_u(g)+ h.c]\notag\\
	&=\frac{1}{N_{\textbf{\textit{d}}}}\frac{1}{|\xi|}\sum_{\textbf{\textit{d}}'}\frac{1}{2N_b}Tr[U_{\varepsilon,l}^{\dagger}(\mathcal{R}_d^{-1}(g)\textbf{\textit{d}}')i\Lambda_nU_{\varepsilon,l'}(\mathcal{R}_d^{-1}(g)\textbf{\textit{d}}')+ h.c]\notag\\
	&=\frac{1}{N_{\textbf{\textit{d}}}}\frac{1}{2N_b}Tr[U_{\varepsilon,l}^{\dagger}(\textbf{\textit{d}}_0)i\Lambda_nU_{\varepsilon,l'}(\textbf{\textit{d}}_0)+ h.c],
\end{align}
which is indeed the same as that in Eq.~\ref{e1}.
In the third line of Eq.~\ref{s3}, we have exploited the following identities:
\begin{align}
	\mathcal{R}_u(g)U_{\varepsilon}(\mathcal{R}_d^{-1}(g)\textbf{\textit{d}})\mathcal{R}_u^\dagger(g)=U_{\varepsilon}(\textbf{\textit{d}}),\notag\\	
	\mathcal{R}_u^\dagger(g)U_{\varepsilon}(\textbf{\textit{d}})\mathcal{R}_u(g)=U_{\varepsilon}(\mathcal{R}_d^{-1}(g)\textbf{\textit{d}}).
\end{align}

\subsection{The details of initial similarity distribution without adiabatic deformation}

In the process of constructing similarity matrices, the most time-consuming step is the adiabatic deformation step. It is natural to see what happens with the similarity distribution without adiabatic deformation.

It is enough to identify phase transition boundaries if we do not care about if two nonadjacent regions belong to the same phase. The phase boundaries can then be detected more easily from the similarity distribution without adiabatic deformation. Here, we offer an intuitive explanation. The geometries of samples in the same region can be twisted slightly into each other, so the similarities for this region are large. For two samples from adjacent regions, the change of their geometries are drastic and the similarity are very small since they have different topological properties. For two samples belonging to the same topological phase but from two nonadjacent regions, their geometries are similar but still need a rotation to change into each other, so their similarity can be very small. Overall, at a phase transition point, an apparent similarity contraction will appear.

To explain this fact, let us use the Floquet SSH model with $\theta=0.75\pi$ and $\gamma \in [-\pi,\pi]$ as the example. The relative similarities for $\pi$ gap and 0 gap without adiabatic deformation is shown in Fig.~\ref{f7}.
\begin{figure}[htbp]
	\centering
	\includegraphics[scale=0.3]{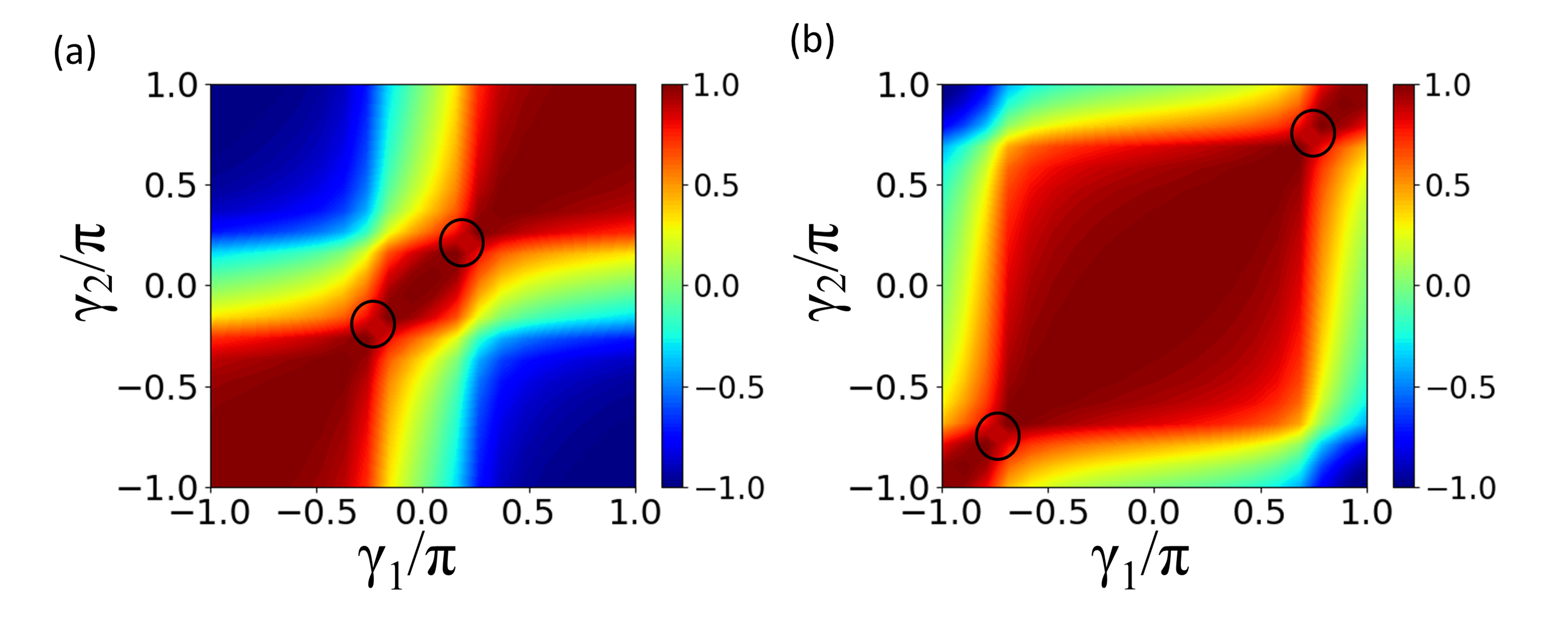}
	\caption{\label{f7} {\bf The results of unprocessed similarities for $\pi$ gap, 0 gap with $ \theta=0.75\pi $ and $ \gamma\in [-\pi,\pi] $}. Black circles mark the contractions of similarity, which signals that the phase transition occurs 
		({\bf a}) Results for the $\pi$ gap. Two circles at $\gamma=\pm \frac{\pi}{4}$ indicate that there are phase transition at these two values.
		({\bf b})Results for the $\pi$ gap. Two circles at $\gamma=\pm \frac{3\pi}{4}$ indicate that there are phase transition at these two values. Here, $\eta=0.01, \epsilon=0.03, f=0.5, N_t=10, N_k=20$.}
\end{figure}

In Fig.~\ref{f7}a(b), the similarities in three regions,$ \gamma\in [-\pi,-\frac{\pi}{4}],[-\frac{\pi}{4},\frac{\pi}{4}]  $ and $[\frac{\pi}{4},\pi]$($\gamma\in [-\pi,-\frac{3\pi}{4}],[-\frac{3\pi}{4},\frac{3\pi}{4}]  $ and $[\frac{3\pi}{4},\pi]$), are almost one and form three red approximate squares. The similarities connecting two side regions( $\left|\gamma\right|>\frac{\pi}{4} $ for $\pi$ gap and $\left|\gamma\right|>\frac{3\pi}{4} $ for 0 gap) with the same topological phase are almost -1, which means that they are not identified to be the same phase in this process(this is why adiabatic deformation are necessary for exploring topological phase completely). These distinct similarity contractions (marked with black circles) indicate the separation of different phases. These signals mark out the phase transitions at $ \gamma=\pm \frac{\pi}{4}$ in $\pi$ gap and $\gamma=\pm \frac{3\pi}{4}$ in 0 gap.  

\end{widetext}

\end{document}